\documentclass[lineno]{biometrika}

\usepackage{amsmath}

\usepackage{times}
\usepackage{bm}
\usepackage{natbib}
\RequirePackage[pdftex]{graphics}
\RequirePackage{graphicx}
\RequirePackage{float}
\usepackage[plain,noend]{algorithm2e}

\makeatletter
\renewcommand{\algocf@captiontext}[2]{#1\algocf@typo. \AlCapFnt{}#2} % text of caption
\def\@algocf@capt@plain{top}
\renewcommand{\algocf@makecaption}[2]{%
  \addtolength{\hsize}{\algomargin}%
  \sbox\@tempboxa{\algocf@captiontext{#1}{#2}}%
  \ifdim\wd\@tempboxa >\hsize%     % if caption is longer than a line
    \hskip .5\algomargin%
    \parbox[t]{\hsize}{\algocf@captiontext{#1}{#2}}% then caption is not centered
  \else%
    \global\@minipagefalse%
    \hbox to\hsize{\box\@tempboxa}% else caption is centered
  \fi%
  \addtolength{\hsize}{-\algomargin}%
}
\makeatother

\begin{document}

%\jname{Biometrika}
%%% The year, volume, and number are determined on publication
%\jyear{2012}
%\jvol{99}
%\jnum{1}
%% The \doi{...} and \accessdate commands are used by the production team
%\doi{10.1093/biomet/asm023}
%\accessdate{Advance Access publication on 31 July 2012}
%\copyrightinfo{\Copyright\ 2012 Biometrika Trust\goodbreak {\em Printed in Great Britain}}

%% These dates are usually set by the production team
%\received{April 2012}
%\revised{September 2012}

%% The left and right page headers are defined here:
\markboth{Dan Shen, Haipeng Shen, Hongtu Zhu \and J.S. Marron}{Miscellanea}

%% Here are the title, author names and addresses
\title{High Dimensional Principal Component Scores and Data Visualization}

%\author{A. C. DAVISON, R. GESSNER}
%\affil{Institute of Mathematics, Ecole Polytechnique F\'ed\'erale de Lausanne, Station 8,\\ 1015 Lausanne, Switzerland \email{editor.biometrika@epfl.ch} \email{biometrika@epfl.ch}}

\author{Dan Shen$^{1}$, Haipeng Shen$^{2}$, Hongtu Zhu$^{1}$, and J. S. Marron$^{1, 2}$}

\affil{$^{1}$Department of Biostatistics, University of North Carolina at Chapel Hill, Chapel Hill, North Carolina 27599, U.S.A.\\\vspace{1mm}
$^{2}$Department of Statistics and Operations Research, University of North Carolina at Chapel Hill, Chapel Hill, North Carolina 27599, U.S.A. \\\vspace{1.8mm}
dshen@email.unc.edu \quad  haipeng@email.unc.edu \\htzhu@email.unc.edu
\quad  marron@email.unc.edu}

\maketitle

\begin{abstract}
Principal component analysis is a useful dimension reduction and data visualization method.
However, in high dimension, low sample size asymptotic contexts,
where the sample size is fixed and the dimension goes to infinity,
a paradox has arisen.
 In particular, despite the useful real data insights commonly obtained from principal component score visualization,
 these scores are not consistent even when the sample eigenvectors are consistent.
 This paradox is resolved by asymptotic study of the ratio between
 the sample and population principal component scores.
 In particular, it is seen that this proportion converges to a non-degenerate  random variable.
 The realization is the same for each data point, i.e. there is a common random rescaling, which appears for each eigen-direction.  This then gives inconsistent axis labels for the standard scores plot, yet the relative positions of the points (typically the main visual content) are consistent.
This paradox disappears when the sample size goes to infinity.
 \end{abstract}

\begin{keywords}
Principal Component Analysis; High Dimension; Low Sample Size; Spike Model.%\vspace{-5mm}
\end{keywords}

\section{Introduction}~\label{introduction}

Visualization of high dimension, low sample size data
\begin{figure}[h]
\vspace{-4.9cm}
 \begin{center}
 \includegraphics[width=\textwidth]{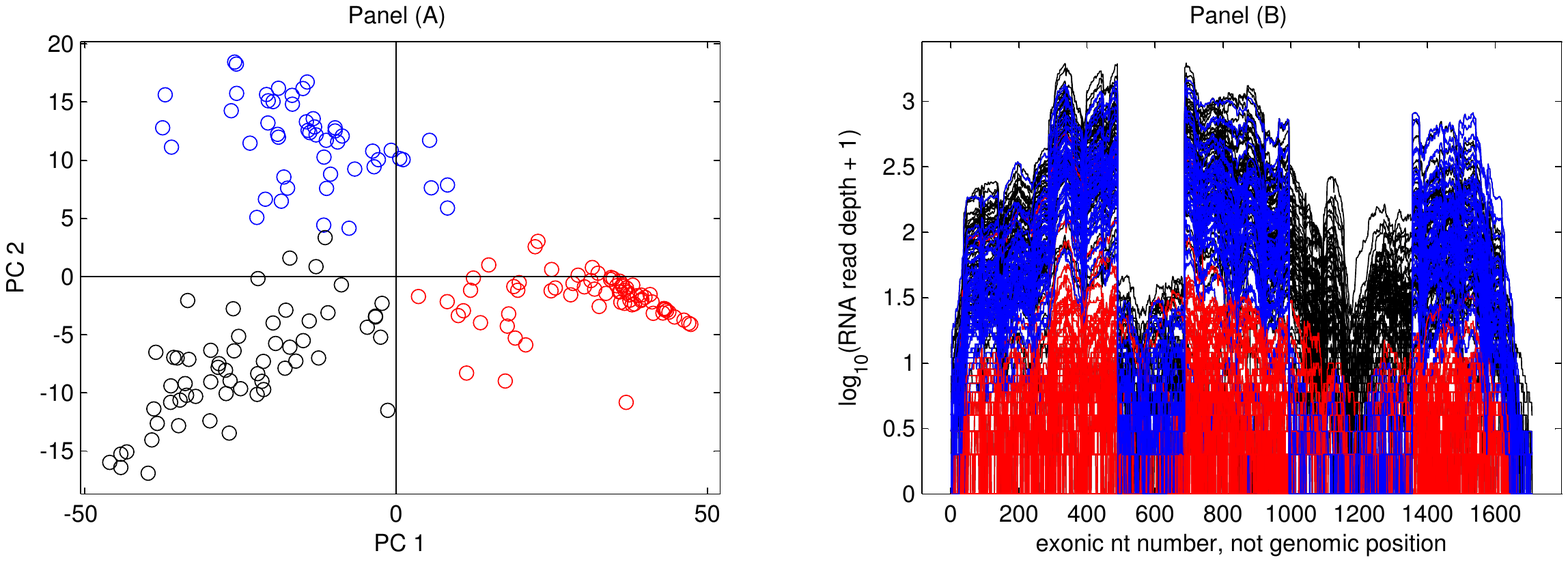}\vspace{-.7cm}
 \end{center}
  \caption{Principal component analysis of a Next Generation Sequencing cancer data set.  The scores plot in Panel (A) suggests three clusters, which are manually brushed with colors.  Relevance of these visually discovered clusters is studied in Panel (B), by showing curves with the brushed colors, which reveals a biologically important alternate splicing event.}\label{fig:01}
 \vspace{-.1cm}
 \end{figure} by principal component analysis
has proven to be very useful. A recent example is shown in Figure~\ref{fig:01},
which studies Next Generation Sequencing for a single gene, in a cancer study, from The Cancer Genome Atlas~\citep{TCGA}.  The data objects are $n = 180$ curves (each from one biological tissue sample), reflecting the log base 10 read depth, at around $d = 1700$ genome map locations. Relative positions of these biological samples are visualized, using a standard principal components scores plot, in Panel (A) of Figure~\ref{fig:01}. The plot shows the projection of the data onto the subspace generated by the first two eigenvectors.  Note that there is distinct visual impression of three clusters.  To investigate this clustering, the clusters have been manually brushed, with three different colors, as shown.  To investigate whether these clusters represent important scientific phenomena, the same coloring is applied to the raw data curves in Panel (B).  The distinct blocks in Panel (B) represent different exonic regions of the genome, and the jumps in the curves at the boundaries of these blocks indicate splicing events.  The red curves are generally very low (recall the log scale) indicating very low levels of expression of this gene, for these samples.  The black and blue curves are generally much higher, showing much stronger gene expression.  The black and blue curve differ strongly over the fourth exonic region, between 1000 and 1400, where the blue samples show a clear deletion of this exon.  Finding such deletions is an important goal in cancer research, as they can form the basis of targeted treatments.  Note that this is just one example, where scientifically important structure in data has been discovered by principal component analysis
in a high dimension, low sample size context.

We are interested in investigating the mathematical underpinnings of this visual approach to data analysis demonstrated in Figure~\ref{fig:01}.  There are several approaches to this in the literature.  Given the nature of genetic data, we prefer to study high dimension, low sample size asymptotics. This approach considers increasing dimension $d\rightarrow\infty$ for a fixed sample size $n$. It has recently been studied in various multivariate analysis contexts, including geometric representation of high dimensional data~\citep{hall2005geometric}, clustering~\citep{Ahn2012Clustering}, and principal component analysis~\citep{ahn2007high, jung2009pca,jung2012boundary,yata2012effective,shen2011}. However, these asymptotic analyses of principal component analysis all focused on studying the angles between the sample eigenvectors and the corresponding population eigenvectors. For example, under some mild conditions,~\cite{jung2009pca} showed that such angles go to 0, which is defined as the \emph{consistency} of the sample eigenvectors.

In this paper, we take a deeper look by studying principal component scores, shown as the circles in Panel (A) of Figure~\ref{fig:01}. Our analysis surprisingly reveals an apparent paradox under the high dimension low sample size setting:  principal component scores are {\it inconsistent} with the corresponding population scores, even when the sample eigenvectors are consistent. Furthermore, for a fixed $n$ and a particular principal component, as $d$ goes to infinity, the proportion between the sample scores and the corresponding population scores converges to a random variable, whose realization is the same for each observation.
The findings suggest that, although the principal component scores can not be consistently estimated, the scores scatter plots, such as Panel (A) of Figure~\ref{fig:01}, can still be used to explore interesting features of high dimension, low sample size data, because the relative positions of the points are consistent due to the common scaling. Finally, this phenomenon disappears when the sample size tends to infinity. In particular, both the sample eigenvectors and the sample principal component scores are then consistent.

\section{Notations and Assumptions}\label{notation}

Assume that $X_1,\ldots, X_n$ are a random sample from the
$d$-dimensional normal distribution $N(\xi, \Sigma)$, and the population covariance
matrix $\Sigma$ has the following eigen-decomposition:
$$
\Sigma=U\Lambda U^T,
$$
where $\Lambda$ is the diagonal matrix of the population eigenvalues
$\lambda_1\geq\ldots\geq\lambda_d$, and $U$
is the corresponding eigenvector matrix such that $U=[u_1,\ldots,u_d]$. Denote the $j$th normalized population
principal component score vector as
\begin{equation}
S_j=(S_{1,j},\cdots, S_{n,j})^T\equiv \lambda_j^{-\frac{1}{2}}(u_j^T X_1,\cdots,  u_j^TX_n)^T, \quad j=1,\cdots, d.
\label{PCs_population}
\end{equation}

Let $\overline{X}$ be the sample mean. As discussed in~\cite{paul2007augmented},
\begin{equation*}
\sum_{i=1}^n (X_i-\overline{X})(X_i-\overline{X})^T \quad \mbox{has the same distribution as} \quad
\sum_{i=1}^{n-1} Y_iY_i^T,
\label{centered-CovarinaceMatrix}
\end{equation*}
where $Y_i$ are independent and identically distributed random variables from $N(0,\Sigma)$.
It follows that the sample covariance matrix is location invariant.
Without loss of generality, we assume that $X_1,\ldots, X_n$ are
a random sample from the $d$-dimensional normal distribution $N(0, \Sigma)$.

Denote the data matrix as $X=[X_1, \ldots, X_n]$, and the sample covariance matrix as $\hat{\Sigma}=n^{-1}XX^{T}$, which has the following eigen-decomposition,
\begin{equation}
\hat{\Sigma}=\hat{U}\hat{\Lambda}\hat{U}^T,
\label{sample:covariance}
\end{equation}
where $\hat{\Lambda}=\mbox{diag} (\hat{\lambda}_1, \cdots, \hat{\lambda}_d)$
is the diagonal sample eigenvalue matrix, and $\hat{U}=[\hat{u}_1,\ldots,\hat{u}_d]$ is
the corresponding sample eigenvector matrix. In addition, the matrix $X/\sqrt{n}$ has the following singular value decomposition: $X/\sqrt{n}=\sum_{j=1}^d \hat{\lambda}_j^{\frac{1}{2}}\hat{u}_j\hat{v}_j$, where $\hat{v}_j=(\hat{v}_{1,j},\cdots,\hat{v}_{n,j})$, $j=1, \cdots, d$. Then the $j$th normalized sample principal component vector is
\begin{equation}
\hat{S}_j=(\hat{S}_{1,j},\cdots, \hat{S}_{n,j})^T=(\hat{v}_{1,j},\cdots,\hat{v}_{n,j})^T, \quad j=1,\cdots, d.
\label{PCs_sample}
\end{equation}
{{Panel (A) of Figure~\ref{fig:01} shows a scatter plot of the $\hat{S}_{i,1}$ versus $\hat{S}_{i,2}$, $i=1,\cdots, n$.}}

\section{Asymptotic Properties of Principal Component Scores}\label{PCs}

The asymptotic properties of principle component scores in high dimension,
low sample size contexts are studied in Section~\ref{PCs_HDLSS}, and as the sample size grows
in Section~\ref{PCs_n}.

\subsection{High Dimension, Low Sample Size Analysis}\label{PCs_HDLSS}

In this subsection, we consider the high dimension, low sample size settings, where
the sample size $n$ is fixed and the dimension $d$ goes to infinity.
We consider multiple spike models~\citep{jung2009pca} under which, as $d \rightarrow\infty$,
\begin{equation}
\lambda_1\gg \cdots \gg \lambda_m \gg \lambda_{m+1}\sim \cdots \sim \lambda_d\sim 1,
\label{multi_HDLSS}
\end{equation}
where $\lambda_i \gg \lambda_j$ means that $\mbox{lim}_{d\rightarrow\infty}\lambda_j/\lambda_i=0$,
and $\lambda_i\sim\lambda_j$ means that $c_1\leq\underline{\mbox{lim}}_{d\rightarrow\infty}\lambda_i/\lambda_j\leq \overline{\mbox{lim}}_{d\rightarrow\infty}\lambda_i/\lambda_j\leq c_2$ for constants $c_1\leq c_2$.

Under the above spike models, \cite{jung2009pca} showed that when $n$ is fixed, if $d/\lambda_m \rightarrow 0$,  the
angle between each of the first $m$ sample eigenvectors $\hat{u}_j$ and its corresponding population eigenvector
$u_j$  goes to 0 with probability 1, which is defined as
 the {\it consistency} of the sample eigenvector.

However, under the same assumptions, an anonymous reviewer  identified a paradoxical phenomenon in that
the sample principal component scores are not consistent. In addition, our analysis suggests that, for a particular principal component,
the proportion between the sample principal component scores and the corresponding population
scores converges to a random variable, the realization of which remains the same for all data points. These results are summarized in the following Theorem~\ref{th:PCs_HDLSS}. The findings suggest that it remains valid to use score scatter plots as a graphical tool to identify interesting features in high dimension low sample size data.

\begin{theorem} Under Assumption~\eqref{multi_HDLSS}, and for the fixed $n$, as $d\rightarrow \infty$,
if $d/\lambda_m \rightarrow 0$, {{then the proportion between the sample and population principal component scores satisfies}}
%then the angle between $\hat{u}_j$ and  $u_j$ converges to 0 with probability 1, but
\begin{equation}
\left|\frac{\hat{S}_{i,j}}{S_{i,j}}\right|\xrightarrow{p} R_j, \quad i=1,\cdots n, \;
j=1,\cdots, m,
\label{PCs_chisqr}
\end{equation}
 where $\xrightarrow{p}$ stands for convergence in probability, and $R_j$ has the same distribution as $\sqrt{n/\chi_n^2}$ with $\chi_n^2$ being the Chi-square distribution with $n$ degrees of freedom.
 \label{th:PCs_HDLSS}
 \end{theorem}

%\textbf{\emph{We now give some remarks about Theorem~\ref{th:PCs_HDLSS}.}}
{{Remark 1.} Under the assumptions in Theorem~\ref{th:PCs_HDLSS},~\cite{jung2009pca} and~\cite{shen2012gPCA} have shown that the angle between the sample eigenvector $\hat{u}_j$
and  the corresponding population eigenvector $u_j$, for $j=1,\cdots, m$, converges to 0 with probability 1, which suggests that the sample eigenvectors are consistent, although the principal component scores are inconsistent under the same assumptions.

{{Remark 2.}  It follows from~\eqref{PCs_chisqr} that the ratio $R_j$ only depends on $j$ (the index of the principal components), but not $i$ (the index of the data points). This particular scaling suggests that the scores scatter plot, such as Panel (A) of Figure~\ref{fig:01}, has incorrectly labeled axes (by the common factor $R_j$ for the corresponding axis), and yet asymptotically correct relative positions of the points; hence the scatter plot still enables meaningful identification of useful scientific features   as demonstrated in Panel (B).}

%From~\eqref{PCs_chisqr}, we find that the proportion between the sample scores and the corresponding population scores converges
%to $R_j$, a random variable that is independent of the sample index $i$.
%In addition, note that $R_j$ converges almost surely to 1 as $n\rightarrow \infty$, which suggests the proportional error
%diminishes as the sample size increases to infinity.

\subsection{Growing Sample Size Analysis}\label{PCs_n}

In this subsection, we consider growing sample size contexts, where $n\rightarrow \infty$,
 and then study the asymptotic properties of the principal component scores.
 This includes a wide range of settings, including classical asymptotics, where  dimension $d$ is fixed, random matrix asymptotics where $d\sim n$ and more, see~\cite{shen2012gPCA} for an overview.
 Unlike the low sample size setting, the apparent inconsistency paradox now disappears. This means
 that both the sample eigenvectors and the sample principal component scores can be consistent.

We consider the following multiple spike models, as $n\rightarrow \infty$,
\begin{equation}
\lambda_1\succ \cdots \succ\lambda_m \gg \lambda_{m+1}\sim \cdots \sim \lambda_d\sim 1.
\label{multi_n}
\end{equation}
Here $\lambda_i\succ\lambda_j$ means that $\overline{\mbox{lim}}_{n\rightarrow\infty}\lambda_j/\lambda_i<1$.
Compared with the multiple spike models~\eqref{multi_HDLSS}, the multiple spike assumption~\eqref{multi_n} is weaker because we have more
sample information ($n\rightarrow \infty$).

%Theorem~\eqref{th:PCs_HDLSS} suggests that the proportional error between the sample scores and the
%corresponding population scores diminishes when the sample size becomes larger and larger.
Theorem~\ref{th:PCs_n} suggests that as $n\rightarrow \infty$, the proportion between the sample scores and the
corresponding population scores tends to 1. This connects with the above results, from the fact that the ratio $R_j$ in~\eqref{PCs_chisqr} has the same distribution as an asymptotic
$\sqrt{n/\chi^2_n}$ distribution which converges almost surely  to 1 as $n\rightarrow\infty$. Thus, it is not surprising that
 the  apparent inconsistency disappears as the sample size grows.

\begin{theorem} Under Assumption~\eqref{multi_n}, and as $n\rightarrow \infty$,
if $d/\lambda_m \rightarrow 0$, {then the proportion between the sample and population principal component scores satisfies}
\begin{equation}
\left|\frac{\hat{S}_{i,j}}{S_{i,j}}\right|\xrightarrow{a.s} 1, \quad i=1,\cdots n, \;
j=1,\cdots, m,
\label{PCs_consistency }
\end{equation}
where $\xrightarrow{a.s}$ stands for almost sure convergence.
 \label{th:PCs_n}
\end{theorem}

{{Remark 1.} Under the current context, the consistency of the sample principal component scores fits as expected, with the fact that the sample eigenvectors are consistent under the assumptions of Theorem~\ref{th:PCs_n}. In particular,~\cite{shen2012gPCA} have shown that, under the same assumptions, the angle between the sample eigenvector $\hat{u}_j$ and the corresponding population eigenvector $u_j$ for $j=1,\cdots,m$ converges almost surely to 0.}

%\appendix

\appendixone
\section*{Appendix}
%\subsection*{Technical details}

In this section, we provide the technical details of the proofs for Theorems~\ref{th:PCs_HDLSS} and~\ref{th:PCs_n}. First, we present two lemmas from~\cite{shen2012gPCA}, that will be used to prove Theorems~\ref{th:PCs_HDLSS} and~\ref{th:PCs_n}.

%\begin{lemma} (Cauchy-Schwarz Inequality). Assuming that $\{a_k, k=1,\cdots, h\}$ and
%$\{b_k, k=1,\cdots, h\}$ are two sequence of real numbers, then we have
%\begin{equation}
%(\sum_{k=1}^h a_k b_k)^2\leq (\sum_{k=1}^h a^2_k)(\sum_{k=1}^h b^2_k).
%\label{CauchySchwarz}
%\end{equation}
%The equation holds if and only if $a_k=b_k$ for $k=1,\cdots,h$.
%\label{lemma:CauchySchwarz}
%\end{lemma}

\begin{lemma}
Under the assumptions in Theorem~\ref{th:PCs_HDLSS} and as $d\rightarrow\infty$, the sample eigenvalues
satisfy
\begin{equation}
\frac{\hat{\lambda}_j}{\lambda_j}\xrightarrow{p} \frac{\chi_n^2}{n}, \quad j=1,\cdots, m,
\label{eigenvalues:hdlss}
\end{equation}
and the sample eigenvectors satisfy
\begin{equation}
\left\{ \begin{array}{ll}
\lambda^{\frac{1}{2}}_k\lambda^{-\frac{1}{2}}_j|\hat{u}_j^T u_k|\xrightarrow{p} 1, \quad \mbox{for} \; 1\leq k=j\leq m, \quad
\mbox{or} \quad 0\;  \;\mbox{for} \;1\leq k\neq j\leq m,\\
 \sum_{k=m+1}^d (\hat{u}_j^T u_k)^2 \xrightarrow{p} 0, \quad \mbox{for}\; 1\leq j\leq m .
 \end{array} \right.
\label{eigenvectors:hdlss}
\end{equation}
\label{lemma:hdlss}
\end{lemma}

\begin{lemma}
Under the assumptions in Theorem~\ref{th:PCs_n} and as $n\rightarrow\infty$, the sample eigenvalues
satisfy
\begin{equation}
\frac{\hat{\lambda}_j}{\lambda_j}\xrightarrow{a.s} 1, \quad j=1,\cdots, m,
\label{eigenvalues:n}
\end{equation}
and the sample eigenvectors satisfy
\begin{equation}
\left\{ \begin{array}{ll}
\lambda^{\frac{1}{2}}_k\lambda^{-\frac{1}{2}}_j|\hat{u}_j^T u_k|\xrightarrow{a.s} 1, \quad \mbox{for} \; 1\leq k=j\leq m, \quad
\mbox{or} \quad 0\;  \;\mbox{for} \;1\leq k\neq j\leq m,\\
 \sum_{k=m+1}^d (\hat{u}_j^T u_k)^2 \xrightarrow{a.s} 0, \quad \mbox{for}\; 1\leq j\leq m .
 \end{array} \right.
\label{eigenvectors:n}
\end{equation}
\label{lemma:n}
\end{lemma}

Note that $X_i$ has the following decomposition
\begin{equation}
X_i=\sum_{j=1}^{d} \lambda_j^{\frac{1}{2}} u_j z_{i,j},
\label{sample_decomposition}
\end{equation}
where the $z_{i,j}$'s are independent and identically standard normally distributed for $i=1,\cdots, n$,
 $j=1,\cdots, d$. It follows from~\eqref{PCs_population} and~\eqref{sample_decomposition}
 that the $j$th population principal component scores are
\begin{equation}
S_j=(S_{1,j},\cdots, S_{n,j})^T=(z_{1,j},\cdots, z_{n,j})^T.
\label{PCs_population_proof}
\end{equation}
From~\eqref{PCs_sample}, the $j$th sample principal component scores are
\begin{equation}
\hat{S}_j=(\hat{S}_{1,j},\cdots, \hat{S}_{n,j})=\hat{\lambda}_j^{-\frac{1}{2}}(\hat{u}_j^T X_1,\cdots,  \hat{u}_j^TX_n).
\label{PCs_sample_proof}
\end{equation}
From~\eqref{sample_decomposition},~\eqref{PCs_population_proof} and~\eqref{PCs_sample_proof},
we have that the proportion between the sample principal component scores and the corresponding population scores
are
\begin{equation}
\frac{\hat{S}_{i,j}}{S_{i,j}}= \frac{\lambda^{\frac{1}{2}}_j}{\hat{\lambda}^{\frac{1}{2}}_j }\hat{u}_j^T u_j
+\sum_{1\leq k \leq m, k\neq j}^{m}\frac{\lambda^{\frac{1}{2}}_k z_{i,k}}{\hat{\lambda}^{\frac{1}{2}}_j z_{i,j}} \hat{u}_j^T u_k
+\sum_{k=m+1}^d \frac{\lambda^{\frac{1}{2}}_k z_{i,k}}{\hat{\lambda}^{\frac{1}{2}}_j z_{i,j}} \hat{u}_j^T u_k.
\label{proportion_PCs}
\end{equation}

\emph{Proof of Theorem~\ref{th:PCs_HDLSS}.} It follows from Lemma~\ref{lemma:hdlss} that as $d\rightarrow \infty$
\begin{equation}
|\frac{\lambda^{\frac{1}{2}}_j}{\hat{\lambda}^{\frac{1}{2}}_j }\hat{u}_j^T u_j|\xrightarrow{p} R_j, \quad
\sum_{1\leq k \leq m, k\neq j}^{m}\frac{\lambda^{\frac{1}{2}}_k z_{i,k}}{\hat{\lambda}^{\frac{1}{2}}_j z_{i,j}} \hat{u}_j^T u_k \xrightarrow{p} 0,
\label{12_estimate}
\end{equation}
where $R_j$ has the same distribution as $\sqrt{n/\chi_n^2}$.
Without loss of generality, we assume that $\lambda_k=1$ for $k=m+1,\cdots, d$. Then it follows from Cauchy-Schwarz inequality that
\begin{equation}
\left\{\sum_{k=m+1}^d \frac{\lambda^{\frac{1}{2}}_k z_{i,k}}{\hat{\lambda}^{\frac{1}{2}}_j z_{i,j}} \hat{u}_j^T u_k\right\}^2 \leq \frac{d-m}{\hat{\lambda}_jz^2_{i,j}}\left\{\frac{1}{d-m}\sum_{k=m+1}^d  z^2_{i,k}\right\}
\left\{\sum_{k=m+1}^d (\hat{u}_j^T u_k)^2\right\}^2.
\label{cauchy_noise}
\end{equation}
 From Lemma~\ref{lemma:hdlss} and~\eqref{cauchy_noise},
  the last item in the right-hand-side of Equation~\eqref{proportion_PCs} converges to 0 with probability 1. Combining the above with~\eqref{12_estimate},
we  obtain~\eqref{PCs_chisqr}. In addition, it follows from~\eqref{eigenvectors:hdlss} that
 the angle between $\hat{u}_j$ and $u_j$ converges to 0 with probability 1, which concludes the proof of
 Theorem~\ref{th:PCs_HDLSS}.

\emph{Proof of Theorem~\ref{th:PCs_n}.} The proof of Theorem~\ref{th:PCs_n} is similar.
To avoid overlap, details are not given here. The critical difference in the proof of Theorem~\ref{th:PCs_n} is the use of Lemma~\ref{lemma:n}, i.e.~\eqref{eigenvalues:hdlss} should be replaced by~\eqref{eigenvalues:n}.

\bibliographystyle{biometrika}
\bibliography{PCScore}

\end{document}